\documentclass[aps,twocolumn]{revtex4}%
\usepackage{amsmath}
\usepackage{graphicx}
\usepackage{epsfig}
\usepackage{amsfonts}
\usepackage{amssymb}%
\begin{document}
\title{The quantum brachistochrone problem for non-Hermitian Hamiltonians}
\author{Paulo E.G. Assis and Andreas Fring}
\affiliation{Centre for Mathematical Science, City University, Northampton Square, London
EC1V 0HB, UK}
\email{Paulo.Goncalves-De-Assis.1@city.ac.uk, A.Fring@city.ac.uk}

\begin{abstract}
Recently Bender, Brody, Jones and Meister found that in the quantum
brachistochrone problem the passage time needed for the evolution of certain
initial states into specified final states can be made arbitrarily small, when
the time-evolution operator is taken to be non-Hermitian but $\mathcal{PT}%
$-symmetric. Here we demonstrate that such phenomena can also be obtained for
non-Hermitian Hamiltonians for which $\mathcal{PT}$-symmetry is completely
broken, i.e. dissipative systems. We observe that the effect of a tunable
passage time can be achieved by projecting between orthogonal eigenstates 
by means of a time-evolution operator
associated to a non-Hermitian Hamiltonian. It is not essential that this
Hamiltonian is $\mathcal{PT}$-symmetric.

\end{abstract}
\maketitle
\date{}
\volumeyear{year}
\volumenumber{number}
\issuenumber{number}
\eid{identifier}
\startpage{1}
\endpage{2}


\section{Introduction}

To find the brachistochrone is one of the oldest problems in classical
mechanics tracing back to Newton and Leibniz. It consists of finding the
trajectory between two locations of a particle, subject to a gravitational
field, for which the transition time becomes minimal. This problem can be
generalized to a relativistic \cite{Goldstein:1985bx} and to a quantum
mechanical setting \cite{Miyake,DB1,Carlini,DB2,faster}. In the latter case
one seeks the minimal time $t=:\tau$, referred to as passage time, such that%
\begin{equation}
\left\vert \psi_{f}\right\rangle =e^{-i\tau H}\left\vert \psi_{i}\right\rangle
, \label{QB}%
\end{equation}
for given initial and final states $|\psi_{i}\rangle$ and $|\psi_{f}\rangle$,
respectively. Equality can be achieved by possibly tuning some parameters in
the Hamiltonian $H$.

Bender, Brody, Jones and Meister \cite{faster} extended this treatment by
allowing also non-Hermitian Hamiltonians in (\ref{QB}). The surprising result
found in \cite{faster} was that when involving non-Hermitian, but
$\mathcal{PT}$-symmetric Hamiltonians, in the evolution operator, the passage
time can be made arbitrarily small by varying a parameter in $H$ while keeping
the transition frequency between two states constant. At present this
phenomenon is an observation and no explanation has been provided as to where
this effect might originate from.

One might suspect that one could make the $\mathcal{PT}$-symmetry responsible
for the observation and seek for similar arguments as those which allow to
explain the reality of the spectrum of a $\mathcal{PT}$-symmetric
non-Hermitian Hamiltonian. See for instance
\cite{special,specialCzech,CArev,Bender:2007nj} for recent results and
reviews. The main purpose of this paper is to investigate whether
$\mathcal{PT}$-symmetry can be utilized as well to explain the observed
phenomenon of a tunable passage time. In fact, we find that the same
conclusion can be drawn when considering non-Hermitian Hamiltonians with
complex eigenvalues describing dissipative systems, i.e. those for which
$\mathcal{PT}$-symmetry is definitely broken. This means the possibility of
arbitrarily small passage times results from the non-Hermitian nature of the
Hamiltonian involved in the time-evolution operator and not its $\mathcal{PT}$-invariance.

Our manuscript is organised as follows: In section II we derive passage times
for $\mathcal{PT}$-symmetric Hamiltonians for various different types of
initial and final states. In section III we perform a similar analysis for
non-Hermitian Hamiltonians with complex eigenvalues deriving similar phenomena
as in section II. We state our conclusions in section IV.

\section{Pseudo Hermitian Hamiltonians}

We start by considering $\mathcal{PT}$-symmetric or rather pseudo Hermitian
Hamiltonians. We recall
\cite{Urubu,Mostafazadeh:2002hb,Mostafazadeh:2001nr,Mostafazadeh:2002id,Mostafazadeh:2003gz}
that a non-Hermitian Hamiltonian $H$ is said to be a pseudo-Hermitian
operator, if there exists a Hermitian operator $\eta$, with regard to the
standard inner product, such that
\begin{equation}
H^{\dagger}=\eta^{2}H\eta^{-2}~~\Leftrightarrow~~h=\eta H\eta^{-1}=h^{\dagger
}. \label{hh}%
\end{equation}
The virtue of such a conjugate pair $h$ and $H$ is that they possess an
identical eigenvalue spectrum, because the Hamiltonians lie in the same
similarity class. The reality of the spectrum is guaranteed, since one of the
Hamiltonians involved, i.e. $h$, is Hermitian. The solutions of the
corresponding time-independent Schr\"{o}dinger equations $H\Phi=\varepsilon
\Phi$ and $h\phi=\varepsilon\phi$ are then simply related as
\begin{equation}
\Phi=\eta^{-1}\phi. \label{23}%
\end{equation}
Let us first discuss the quantum brachistochrone problem for these types of systems.

\subsection{$\phi$ $\rightarrow$ $\phi$ via u $\equiv$ $\Phi\rightarrow\Phi$
via U}

Taking the initial state $\left\vert \psi_{i}\right\rangle $, the final state
$\left\vert \psi_{f}\right\rangle $ to be orthonormal states of a Hermitian
Hamiltonian system and the time-evolution operator in (\ref{QB}) to be
Hermitian as well is the simplest situation to investigate. Equivalent would
be to investigate the non-Hermitian system obtained by the similarity
transformations (\ref{hh}) and (\ref{23}). Here we want to solve the quantum
brachistochrone problem in a slightly less stringent form as studied in
\cite{Miyake,DB1,Carlini,DB2,faster}. Instead of solving (\ref{QB}), we just
consider the physical relevant matrix element and seek the minimal time
$t=:\tau$, such that a given transition probability is reached. This means for
normalized initial and final states we solve the equation%
\begin{equation}
\left\vert \left\langle \phi_{f}\right.  \!\left\vert u(\tau,0)\phi
_{i}\right\rangle \right\vert =\left\vert \left\langle \Phi_{f}\right.
\!\left\vert U(\tau,0)\Phi_{i}\right\rangle _{\eta}\right\vert =\beta,
\label{qb}%
\end{equation}
for $\tau$ with given constant $0\leq\beta\leq1$. Here $u(t,t^{\prime})$ and
$U(t,t^{\prime})=\eta^{-1}u(t,t^{\prime})\eta$ are the time-evolution
operators, which evolve a wavefunction from time $t^{\prime}$ to $t$
associated to the Hermitian and non-Hermitian system, respectively. In order
to make the inner product involving the $\Phi$s meaningful, we have to change
the metric in the second expression in (\ref{qb}) as was argued in
\cite{Urubu,Mostafazadeh:2001nr,Bender:2002vv} or equivalently use a
biorthonormal basis, see e.g. \cite{Ingrid,Rot2}. Here we use the convention
$\langle\psi_{f}\!|\mathcal{O}\psi_{i}\rangle_{\eta}:=\langle\psi_{f}%
\!|\eta^{2}\mathcal{O}\psi_{i}\rangle$ for the $\eta$-inner product with
$\mathcal{O}$ being some operator. Clearly, for given time-evolution operators
and final and initial states a real solution for $\tau$ for all values of
$\beta$ does not always exist. Natural choices are for instance $\beta=1$ or
the maximum transition amplitude.

Starting with time independent Hamiltonians, the problem is solved in a
straightforward manner as we simply have $u(\tau,0)=e^{-i\tau h}$ and
$U(\tau,0)=e^{-i\tau H}$. When $\phi_{i}$ and $\phi_{f}$ are orthogonal states
we can find a solution in complete generality. Taking $|\phi_{+}\rangle$ and
$|\phi_{-}\rangle$ to be two normalized eigenstates of $h$, the two states%
\begin{equation}
\left\vert \phi_{f/i}\right\rangle =\frac{1}{\sqrt{2}}\left(  \left\vert
\phi_{-}\right\rangle \pm i\left\vert \phi_{+}\right\rangle \right)  .
\label{fi}%
\end{equation}
are orthonormal with regard to the standard inner product. It is then
straightforward to compute the matrix element occurring in (\ref{qb})
\begin{align}
\left\langle \phi_{f}\right.  \!\left\vert e^{-ith}\phi_{i}\right\rangle  &
=\frac{1}{\sqrt{2}}\left\langle \phi_{f}\right.  \!\left(  \left\vert
e^{-it\varepsilon_{-}}\phi_{-}\right\rangle -i\!\left\vert e^{-it\varepsilon
_{+}}\phi_{+}\right\rangle \right) \nonumber\\
&  =\frac{1}{2}e^{-i\varepsilon_{-}t}\left(  1-e^{-i\omega t}\right)  ,
\label{a}%
\end{align}
where the transition frequency between the two states is denoted as
$\omega=\varepsilon_{+}-\varepsilon_{-}$. This means the quantum
brachistochrone problem in the version (\ref{qb}) is solved for the passage
time
\begin{equation}
\tau=\frac{2}{\omega}\arcsin\beta. \label{tb}%
\end{equation}
For $\beta=1$ we recover $\tau=\pi/\omega$, which in slightly different forms,
is well known and holds for any Hermitian or equivalent non-Hermitian system,
as specified in (\ref{hh}). As pointed out first in \cite{faster} more
spectacular results can be obtained when involving non-Hermitian Hamiltonians
in the evolution operator while keeping the eigenstates to be associated to a
Hermitian system.

\subsection{$\phi$ $\rightarrow$ $\phi$ via U $\equiv$ $\Phi\rightarrow\Phi$
via u}

In \cite{faster} a Gedankenexperiment was proposed in which a particle passes
through a region, which causes its governing Hamiltonian to change from a
Hermitian to a non-Hermitian one. This scenario implies that the Hamiltonian
becomes explicitly time-dependent. The situation considered in \cite{faster}
was for the initial and final states to be orthogonal states in a Hermitian
system, whereas the time evolution was associated to a non-Hermitian
Hamiltonian. In general, we can write this temporary change of the Hamiltonian
to a non-Hermitian Hamiltonian in the form%
\begin{equation}
H(t)=h+gh_{1}(t), \label{h1}%
\end{equation}
where $h=h^{\dag}$ and $h_{1}\neq h_{1}^{\dag}$. This means we consider an
analogue to a conventional time dependent scenario, however, with the
difference that the perturbation is now non-Hermitian. A standard example of
$H(t)$ in (\ref{h1}) with $h_{1}$ being Hermitian is for instance the
Stark-LoSurdo Hamiltonian describing an atom in an external electric field,
with $h$ representing the unperturbed atomic system and $h_{1}(t)$ the
external electric field. In \cite{CA,CArev} the alternative scenario was
considered in which also the unperturbed system was taken to be non-Hermitian.
The treatment in \cite{faster} corresponds to the special case of (\ref{h1})
in which the time dependence is of the form of a stepfunction. This means to
describe that setting one has to take $H(t)=H=h+gh_{1}$ for $0\leq t\leq\tau$
and $H(t)=h$ for $t>\tau$, with $\tau$ being the passage time rather than the
pulse length as in the aforementioned example.

Then, depending on the choice of the initial and final states, the
time-evolution operator and the inner product, the quantum brachistochrone
problem can be formulated in various different ways from (\ref{qb}). For
instance, when projecting between orthogonal Hermitian states via a
non-Hermitian time-evolution operator one may consider
\begin{equation}
\frac{\left\vert \left\langle \phi_{f}\right.  \!\left\vert U(\tau,0)\phi
_{i}\right\rangle _{\eta}\right\vert }{\left\Vert \phi_{f}\right\Vert _{\eta
}\left\Vert \phi_{i}\right\Vert _{\eta}}=\frac{\left\vert \left\langle
\eta\phi_{f}\right.  \!\left\vert u(\tau,0)\eta\phi_{i}\right\rangle
\right\vert }{\left\vert \eta\phi_{f}\right\vert \left\vert \eta\phi
_{i}\right\vert }=\beta, \label{BC2}%
\end{equation}
where we used the $\eta$-norm defined as $\left\Vert \phi\right\Vert _{\eta
}:=\left\vert \eta\phi\right\vert =\sqrt{\left\langle \eta\phi\right.
\!\left\vert \eta\phi\right\rangle }$. This switching between the eigenstates
is in fact the key point of the entire analysis. We may view equation
(\ref{BC2}) in two equivalent ways. On one hand on the right hand side we just project
with a standard Hermitian time-evolution operator, between two somewhat
unusual, but perfectly viable initial and final states $\eta\phi_{i}$ and
$\eta\phi_{f}$, respectively. This is a picture entirely in the standard
quantum mechanical description with the difference that the initial and final
states are not taken to be orthogonal. On the other hand we may use the equality and
view this expression as a projection of some orthogonal initial and final
states by means of a non-Hermitian time-evolution operator. We stress the fact
that the initial and final states do not propagate in time and that the metric
is not changed in this procedure \cite{note}.

Slightly different and less natural is the possibility corresponding to
(\ref{QB}) when projecting with the standard inner product onto a final state.
When written as an expectation value this amounts to
\begin{equation}
\frac{\left\vert \left\langle \phi_{f}\right.  \!\left\vert U(\tau,0)\phi
_{i}\right\rangle \right\vert }{\left\Vert \phi_{f}\right\Vert _{\eta^{-1}%
}\left\Vert \phi_{i}\right\Vert _{\eta}}=\frac{\left\vert \left\langle
\eta\phi_{f}\right.  \!\left\vert u(\tau,0)\eta\phi_{i}\right\rangle
_{\eta^{-1}}\right\vert }{\left\vert \eta^{-1}\phi_{f}\right\vert \left\vert
\eta\phi_{i}\right\vert }=\beta.\label{BC}%
\end{equation}
Alternatively, one may also envisage a situation when one projects from
orthogonal Hermitian states onto eigenstates of a non-Hermitian Hamiltonian
via a non-Hermitian time-evolution operator, or vice versa. Then one should
solve%
\begin{equation}
\frac{\left\vert \left\langle \Phi_{f}\right.  \!\left\vert U(\tau,0)\phi
_{i}\right\rangle _{\eta}\right\vert }{\left\Vert \Phi_{f}\right\Vert _{\eta
}\left\Vert \phi_{i}\right\Vert _{\eta}}=\frac{\left\vert \left\langle
\phi_{f}\right.  \!\left\vert u(\tau,0)\eta\phi_{i}\right\rangle \right\vert
}{\left\vert \phi_{f}\right\vert \left\vert \eta\phi_{i}\right\vert }%
=\beta,\label{df}%
\end{equation}
or%
\begin{equation}
\frac{\left\vert \left\langle \phi_{f}\right.  \!\left\vert U(\tau,0)\Phi
_{i}\right\rangle _{\eta}\right\vert }{\left\Vert \phi_{f}\right\Vert _{\eta
}\left\Vert \Phi_{i}\right\Vert _{\eta}}=\frac{\left\vert \left\langle
\eta\phi_{f}\right.  \!\left\vert u(\tau,0)\phi_{i}\right\rangle \right\vert
}{\left\vert \eta\phi_{f}\right\vert \left\vert \phi_{i}\right\vert }%
=\beta.\label{df2}%
\end{equation}
As in a conventional time-dependent scenarios one is rarely able to compute
the time-evolution operator exactly. However, assuming the non-Hermitian term
in (\ref{h1}) to be small when compared with $h$, we may apply standard
perturbation theory by iterating the DuHamel formula \cite{AC1,AC2,AC3}
\[
U_{H}(t,t^{\prime})=U_{h}(t,t^{\prime})-ig\int\nolimits_{t^{\prime}}%
^{t}dsU_{H}(t,s)h_{1}(s)U_{h}(s,t^{\prime}).
\]
Taking $U_{h}(t,t^{\prime})=\exp(-ih(t-t^{\prime}))$ we obtain to first order
in $g$%
\begin{equation}
U_{H}(t,0)=e^{-iht}-ig\int\nolimits_{0}^{t}ds~e^{-ih(t-s)}h_{1}(s)e^{-ihs}%
.\label{12}%
\end{equation}
We may then compute for instance perturbatively the matrix element%
\[
\left\langle \phi_{f}\right.  \!\left\vert U_{H}(t,0)\phi_{i}\right\rangle
=\left(  1-e^{-i\omega t}\right)  \left[  \frac{1}{2}+ig\left\langle \phi
_{+}\right.  \!\left\vert \operatorname{Im}h_{1}\phi_{-}\right\rangle \right]
.
\]
When $h_{1}$ is Hermitian we naturally recover the result in (\ref{a}). One
may now proceed perturbatively using the above expression for $U_{H}(t,0)$.
However, it is clear from the previous discussion that essentially all aspects
of the problem, which we wish to consider here, may be illustrated by
selecting a two-level system from the larger, possibly even infinite,
spectrum. Thus without loss of generality one may consider a $2\times2$ matrix Hamiltonian.

In order to set the scene for the next section let us briefly recall with some
minor variation the analysis of \cite{faster}.

\subsection{A $2\times2$ matrix Hamiltonian}

A pair of $2\times2$ matrix Hamiltonians related by a similarity
transformation as in (\ref{hh}) is
\begin{equation}
H=\left(
\begin{array}
[c]{cc}%
re^{i\theta} & s\\
s & re^{-i\theta}%
\end{array}
\right)  ,\text{\quad}h=\left(
\begin{array}
[c]{cc}%
r\cos\theta & -\frac{\omega}{2}\\
-\frac{\omega}{2} & r\cos\theta
\end{array}
\right)  , \label{H}%
\end{equation}
with $\omega=2\sqrt{s^{2}-r^{2}\sin^{2}\theta}$ and $r,s,\theta\in\mathbb{R}$.
For the eigenvalues $\varepsilon_{\pm}=r\cos\theta\pm\omega/2$ to be real, one
requires $s^{2}\geq r^{2}\sin^{2}\theta$, such that it is meaningful to
introduce the new parameterization $\sin\alpha=r/s\sin\theta$ with $\alpha
\in\mathbb{R}$. This parameter range guarantees therefore unbroken
$\mathcal{PT}$-symmetry. The Hamiltonians $h$ and $H$ are related by the
similarity transformation in (\ref{hh}), involving the Hermitian operator
\begin{equation}
\eta=\frac{1}{\sqrt{\cos\alpha}}\left(
\begin{array}
[c]{cc}%
\sin\alpha/2 & -i\cos\alpha/2\\
i\cos\alpha/2 & \sin\alpha/2
\end{array}
\right)  . \label{eta}%
\end{equation}
The normalized eigenstates of $H$ and $h$ are
\begin{equation}
\left\vert \Phi_{\pm}\right\rangle _{\alpha}\!=\!\frac{e^{\frac{i\pi}{4}%
(1\mp1)}}{\sqrt{2\cos\alpha}}\left(
\begin{array}
[c]{c}%
-e^{\pm\frac{i\alpha}{2}}\\
\mp e^{\mp\frac{i\alpha}{2}}%
\end{array}
\right)  ,\left\vert \phi_{\pm}\right\rangle =\frac{e^{\frac{i\pi}{4}(1\pm1)}%
}{\sqrt{2}}\left(
\begin{array}
[c]{c}%
1\\
\mp1
\end{array}
\right)  , \label{ff}%
\end{equation}
respectively. Taking now as initial and final states the orthogonal states
$\left\vert \phi_{i}\right\rangle $ and $\left\vert \phi_{f}\right\rangle $ as
defined in (\ref{fi}), we compute
\begin{equation}
e^{-ith}\left\vert \phi_{i}\right\rangle =e^{-ir\cos\theta t}\binom{\cos
\frac{\omega t}{2}}{i\sin\frac{\omega t}{2}},
\end{equation}
and recover from (\ref{qb}), with $\beta=1$, the passage time $\tau=\pi
/\omega$. This is what we expect from the general expression (\ref{tb}) and in
fact it is the same expression as obtained in \cite{faster}, where essentially
the second equation in (\ref{qb}) was evaluated. On the other hand, if we now
let the particle pass through the region in which the corresponding
Hamiltonian becomes non-Hermitian, we may compute $\tau$ by analyzing
(\ref{BC2}) or possibly (\ref{BC}). Acting with $u(t,0)=e^{-ith}$ on the
transformed initial state yields%
\begin{equation}
e^{-ith}\eta\left\vert \phi_{i}\right\rangle =\frac{e^{-itr\cos\theta}}%
{\sqrt{\cos\alpha}}\binom{\sin\frac{1}{2}(\alpha-t\omega)}{i\cos\frac{1}%
{2}(\alpha-t\omega)}. \label{2}%
\end{equation}
When not acting on eigenstates with the operator $e^{-ith}$ or $e^{-itH}$ one
has to turn the infinite sum of operators into a matrix multiplication, see
e.g. \cite{faster}. For this one can exploit the fact that any $2\times
2$-matrix $M$ can be decomposed in terms of Pauli matrices as $M=\mu
_{0}\mathbb{I+}\mathbf{\mu}\cdot\mathbf{\sigma}$ with $\mu_{i}\in\mathbb{C}$,
$i=0,1,2,3.$ Having expressed $h$ or $H$ in this manner, the operation with
$e^{-ith}$ or $e^{-itH}$ on a state reduces to a simple matrix multiplication
by using the identity $e^{\varphi\mathbf{\mu}\cdot\mathbf{\sigma}}=\cos
\varphi\mathbb{I}+i\sin\varphi\mathbf{\mu}\cdot\mathbf{\sigma}$.

Using (\ref{2}), the matrix element in (\ref{BC2}) is computed to
\begin{equation}
\left\langle \phi_{f}\right.  \!\left\vert U(\tau,0)\phi_{i}\right\rangle
_{\eta}=ie^{-i\tau r\cos\theta}\frac{\sin(\alpha-\frac{\tau\omega}{2})}%
{\cos\alpha}. \label{xx}%
\end{equation}
Choosing the constant $\beta=1$ and substituting (\ref{xx}) into (\ref{BC2}),
we compute with $\left\Vert \phi_{f}\right\Vert _{\eta}\left\Vert \phi
_{i}\right\Vert _{\eta}=1/\cos\alpha$ the passage time $\tau=\pi
/\omega+2\alpha/\omega$. This expression involves now the parameter $\alpha$,
which may be tuned to make $\tau$ arbitrarily small while keeping the
transition frequency constant, as was first pointed out in \cite{faster}.

There are two equivalent ways of looking this result. On one hand we may think
that one has solved the quantum brachistochrone problem entirely within the
framework of the Hermitian system for some states, which have no obvious
intrinsic meaning without referring to the non-Hermitian counterpart. On the
other hand we may think that one has solved the time-dependent problem as
outlined in the previous subsection involving the evolution with a
non-Hermitian Hamiltonian between two orthonomal states in the Hermitian system.

Similarly, we may consider a situation in which the final state is constructed
from eigenstates of the non-Hermitian system and compute instead%
\begin{equation}
\left\langle \Phi_{f}\right.  \!\left\vert U(\tau,0)\phi_{i}\right\rangle
_{\eta}=ie^{-i\tau r\cos\theta}\frac{\cos\frac{1}{2}(\alpha-\tau\omega)}%
{\sqrt{\cos\alpha}}.
\end{equation}
This yields for the same choice of $\beta=1$ the passage time $\tau
=2\pi/\omega+\alpha/\omega$ with $\left\Vert \Phi_{f}\right\Vert _{\eta
}\left\Vert \phi_{i}\right\Vert _{\eta}=1/\sqrt{\cos\alpha}$. We summarize our
findings for the different types of scenarios in the following table:

\begin{center}%
\begin{tabular}
[c]{||l|l|l|l||}\hline
$\left\vert \psi_{i}\right\rangle $ & $\left\vert \psi_{f}\right\rangle $ &
$U(\tau,0)$ & $\tau$\\\hline\hline
$\left\vert \Phi_{i}\right\rangle $ & $\left\vert \Phi_{f}\right\rangle $ &
$e^{-iH\tau}$ & $\pi/\omega$\\\hline
$\left\vert \phi_{i}\right\rangle $ & $\left\vert \phi_{f}\right\rangle $ &
$e^{-iH\tau}$ & $\pi/\omega+2\alpha/\omega$\\\hline
$\left\vert \phi_{i}\right\rangle $ & $\left\vert \Phi_{f}\right\rangle $ &
$e^{-iH\tau}$ & $2\pi/\omega+\alpha/\omega$\\\hline
$\left\vert \Phi_{i}\right\rangle $ & $\left\vert \phi_{f}\right\rangle $ &
$e^{-iH\tau}$ & $2\pi/\omega+\alpha/\omega$\\\hline
\end{tabular}

\end{center}

\noindent{\small {Table 1: Passage times for various different initial and
final states evolved by means of a }}$PT${\small -symmetric {non-Hermitian
time evolution operator. All matrix elements are computed with the $\eta
$-inner product and states are normalized with the $\eta$-norm.}}

Next we demonstrate that such type of behaviour is not limited to
$\mathcal{PT}$-symmetric Hamiltonians, but can also be found for dissipative
systems, i.e. genuinely non-Hermitian Hamiltonians with complex eigenvalues
with negative imaginary part.

\section{Non-Hermitian dissipative Hamiltonian systems}

As argued above it is sufficient to consider a $2\times2$ matrix Hamiltonian.
We will now consider two different types of dissipative systems, i.e. those
which have real and those with complex transition frequencies.

\subsection{Real transition frequency}

Let us modify the Hamiltonian $H$ in (\ref{H}) slightly, such that it becomes
a genuinely dissipative system. In order to achieve this we need to break the
$\mathcal{PT}$-symmetry not only for the wavefunction, but also for the
Hamiltonian. Such type of \ Hamiltonians result for instance as effective
Hamiltonians by coupling two non-degenerate states to some open channel as for
instance explained in \cite{FW,HF}. We consider here such type of Hamiltonian of the 
particular form
\begin{equation}
\tilde{H}=\left(
\begin{array}
[c]{cc}%
E+\varepsilon & 0\\
0 & E-\varepsilon
\end{array}
\right)  -i\lambda\left(
\begin{array}
[c]{cc}%
re^{i\theta} & s\\
s & re^{-i\theta}%
\end{array}
\right)  ,~~ \label{H2}%
\end{equation}
with $E,\varepsilon,r,s,\theta,\lambda\in\mathbb{R}$. Similarly as for
(\ref{H}) we will not provide here a concrete physical meaning for the
parameters, as we would like to keep our treatment as generic as possible.
Note that this Hamiltonian does not simply correspond to going to the regime
of broken $\mathcal{PT}$-symmetry for the Hamiltonian (\ref{H}) of the
previous section. Instead, in the simultaneous limit $E,\varepsilon
\rightarrow0$ and $\lambda\rightarrow i$, the dissipative Hamiltonian system
$\tilde{H}$ reduces to the $\mathcal{PT}$-symmetric Hamiltonian $H$.

The eigenvalues of $\tilde{H}$ are computed to%
\begin{equation}
\tilde{\varepsilon}_{\pm}=E\pm\frac{\tilde{\omega}}{2}-ir\lambda\cos\theta,
\label{et}%
\end{equation}
with $\tilde{\omega}=\tilde{\varepsilon}_{+}-\tilde{\varepsilon}_{-}%
=2\sqrt{(\varepsilon+r\lambda\sin\theta)^{2}-\lambda^{2}s^{2}}$ denoting the
transition frequency. For $\lambda$ being restricted to the interval
$-\varepsilon/(s+\sin\theta)\leq\lambda\leq\varepsilon/(s-\sin\theta)$, we can
guarantee that $\tilde{\omega}\in\mathbb{R}$. In this parameter range the
complex energy eigenvalues are indeed of the desired form of a decaying state,
that is $\tilde{\varepsilon}_{\pm}=E_{\pm}-i\Gamma/2$ with decay width
$\Gamma=2r\lambda\cos\theta\in\mathbb{R}^{+}$ when $-\pi/2\leq\theta\leq\pi
/2$. It is then useful to introduce the parameterization%
\begin{equation}
\sin\tilde{\alpha}=\frac{s\lambda}{\varepsilon+r\lambda\sin\theta},
\end{equation}
such that $\tan\tilde{\alpha}=2s\lambda/\tilde{\omega}$. The right
eigenvectors of $\tilde{H}$ corresponding to the eigenvalues $\tilde
{\varepsilon}_{\pm}$ in (\ref{et}) may then be expressed as
\begin{equation}
\left\vert \tilde{\Phi}_{\pm}\right\rangle =\frac{1}{\sqrt{2}}\left(
\left\vert \Phi_{-}\right\rangle _{\tilde{\alpha}}\pm i\left\vert \Phi
_{+}\right\rangle _{\tilde{\alpha}}\right)  , \label{ev}%
\end{equation}
where $\left\vert \Phi_{\pm}\right\rangle _{\alpha}$ is defined in equation
(\ref{ff}). From the considerations in the previous section it is clear that
the operator $\eta$ is vital for the computations of the matrix elements
occurring in the quantum brachistochrone problem, especially when one wishes
to evolve eigenstates of a Hermitian Hamiltonian with a time-evolution
operator associated to a non-Hermitian system. However, since for the case at
hand the Hamiltonian $\tilde{H}$ is now genuinely complex there can not exist
any similarity transformation, which relates it to a Hermitian Hamiltonian.
Nonetheless, we can use the other property of $\eta$, namely that it can be
utilized to introduce a physically well defined inner product. This means we
can seek a transformation such that the eigenstates (\ref{ev}) become
orthonormal with regard to this product. From (\ref{ev}) it is clear that we
can take the same form for $\eta$, but only have to replace $\alpha$ by
$\tilde{\alpha}$ to define a new $\tilde{\eta}$. With the help of this new
operator we construct the eigenstates $|\tilde{\phi}_{\pm}\rangle=\tilde{\eta
}|\tilde{\Phi}_{\pm}\rangle$, which yield indeed the desired orthogonality
relations
\begin{equation}
\left\langle \tilde{\Phi}_{n}\right.  \!\left\vert \tilde{\Phi}_{m}%
\right\rangle _{\tilde{\eta}}=\left\langle \tilde{\eta}^{-1}\tilde{\phi}%
_{n}\right.  \!\left\vert \tilde{\eta}^{-1}\tilde{\phi}_{m}\right\rangle
_{\tilde{\eta}}=\left\langle \tilde{\phi}_{n}\right.  \!\left\vert \tilde
{\phi}_{m}\right\rangle =\delta_{nm},\text{\quad} \label{ortho}%
\end{equation}
for $n,m\in\{+,-\}$. The states $\tilde{\phi}$ are eigenstates to the analogue
of the Hermitian counterpart of a pseudo-Hermitian Hamiltonian. In fact, the
adjoint action of $\tilde{\eta}$ diagonalizes $\tilde{H}$ as
\begin{equation}
\tilde{h}=\tilde{\eta}\tilde{H}\tilde{\eta}^{-1}=\left(
\begin{array}
[c]{cc}%
E-i\frac{\Gamma}{2}-\frac{\tilde{\omega}}{2} & 0\\
0 & E-i\frac{\Gamma}{2}+\frac{\tilde{\omega}}{2}%
\end{array}
\right)  .
\end{equation}
Obviously we have now $\tilde{h}\neq\tilde{h}^{\dagger}$, but $\tilde{h}$ has
the same eigenvalues as $\tilde{H}$, because it lies in the same similarity class.

\subsubsection{$\tilde{\Phi}\rightarrow\tilde{\Phi}$ via \~{U}}

We are now in the position to solve the quantum brachistochrone problem for
dissipative non-Hermitian Hamiltonians. For this we note first that when we
try to compute the passage time $\tau$ directly from the relation (\ref{qb})
with $\beta=1$, it will turn out to be complex. This would of course always be
the case when the transition amplitude for all real values of $t$ is smaller
than $\beta$. Therefore in order to find a physical solution we need to
reformulate the quantum brachistochrone problem slightly to accommodate also
the dissipative non-Hermitian Hamiltonians. A natural expression to consider
is one which features explicitly the decay width $\Gamma$, such as
\begin{equation}
\left\vert \left\langle \tilde{\phi}_{f}\right.  \!\left\vert \tilde{u}%
(\tau,0)\tilde{\phi}_{i}\right\rangle \right\vert =\left\vert \left\langle
\tilde{\Phi}_{f}\right.  \!\left\vert \tilde{U}(\tau,0)\tilde{\Phi}%
_{i}\right\rangle _{\tilde{\eta}}\right\vert =\tilde{\beta}e^{-\frac{\Gamma
\pi}{2\tilde{\omega}}}, \label{qbnH}%
\end{equation}
for normalized initial and final states with $0\leq\tilde{\beta}\leq1$. This
means for stable particles, i.e. $\Gamma\rightarrow0$, we recover the
expression (\ref{qb}). In analogy to the normalized eigenstates of the
Hermitian Hamiltonian $h$ in (\ref{fi}), we take now the initial and final
states to be
\begin{equation}
\left\vert \tilde{\phi}_{f/i}\right\rangle =\frac{\tilde{\eta}}{\sqrt{2}%
}\left(  \left\vert \tilde{\Phi}_{-}\right\rangle \pm i\left\vert \tilde{\Phi
}_{+}\right\rangle \right)  =\frac{1}{\sqrt{2}}\left(  \left\vert \tilde{\phi
}_{-}\right\rangle \pm i\left\vert \tilde{\phi}_{+}\right\rangle \right)  .
\label{28}%
\end{equation}
Assuming at first no time dependence in the Hamiltonian, we compute
\begin{equation}
\left\langle \tilde{\phi}_{f}\right.  \!\left\vert e^{-it\tilde{h}}\tilde
{\phi}_{i}\right\rangle =\frac{1}{2}e^{-i\tilde{\varepsilon}_{-}t}\left(
1-e^{-i\tilde{\omega}t}\right)  . \label{phit}%
\end{equation}
Solving then (\ref{qbnH}) for the passage time in complete analogy to
(\ref{qb}) with $\tilde{\beta}=1$ yields the same expression for the passage
time, namely $\tau=\pi/\tilde{\omega}$. Alternatively we could have also
computed the second expression in (\ref{qbnH}) using $\tilde{U}%
(t,0)=e^{-it\tilde{H}}$, which would of course lead to the same result.
Possibly more interesting passage times can be obtained when we evolve the
states (\ref{28}) with the analogue to the non-Hermitian time evolution.

\subsubsection{$\tilde{\Phi}\rightarrow\tilde{\Phi}$ via \~{u}}

We may assume now an explicit time dependence as in section II C and try to
evolve the states $|\tilde{\phi}_{i}\rangle$ by means of a time-evolution
operator involving the Hamiltonians $\tilde{H}$ in (\ref{H2}). For that
situation to make sense we need to consider the $\tilde{\eta}$-inner products
and generalize (\ref{BC2}) to
\begin{equation}
\frac{\left\vert \left\langle \tilde{\phi}_{f}\right.  \!\left\vert \tilde
{U}(\tau,0)\tilde{\phi}_{i}\right\rangle _{\tilde{\eta}}\right\vert
}{\left\Vert \tilde{\phi}_{f}\right\Vert _{\tilde{\eta}}\left\Vert \tilde
{\phi}_{i}\right\Vert _{\tilde{\eta}}}=\frac{\left\vert \left\langle
\tilde{\eta}\tilde{\phi}_{f}\right.  \!\left\vert \tilde{u}(\tau,0)\tilde
{\eta}\tilde{\phi}_{i}\right\rangle \right\vert }{\left\vert \tilde{\eta
}\tilde{\phi}_{f}\right\vert \left\vert \tilde{\eta}\tilde{\phi}%
_{i}\right\vert }=\tilde{\beta}e^{-\frac{\Gamma\pi}{2\tilde{\omega}}},
\label{gh}%
\end{equation}
by introducing the decay width on the right hand side, similarly as we
extended (\ref{qb}) to (\ref{qbnH}). This way the passage time will result to
be real. Assuming a simple stepfunction time-dependence in (\ref{h1}), we
compute
\begin{equation}
e^{-it\tilde{h}}\tilde{\eta}\left\vert \tilde{\phi}_{i}\right\rangle
=\frac{e^{-it(E-i\lambda r\cos\theta)}\sqrt{\cos\tilde{\alpha}}}{\sqrt{2}%
(\cos\frac{\tilde{\alpha}}{2}+\sin\frac{\tilde{\alpha}}{2})}\binom
{-e^{it\tilde{\omega}/2}}{ie^{-it\tilde{\omega}/2}}%
\end{equation}
and the matrix element
\begin{align}
\left\langle \tilde{\phi}_{f}\right.  \!\left\vert e^{-it\tilde{H}}\tilde
{\phi}_{i}\right\rangle _{\tilde{\eta}}  &  =\left\langle \tilde{\eta}%
\tilde{\phi}_{f}\right.  \!\left\vert e^{-it\tilde{h}}\tilde{\eta}\tilde{\phi
}_{i}\right\rangle \\
&  =-ie^{-it(E-i\lambda r\cos\theta)}\sin\frac{t\tilde{\omega}}{2}.
\end{align}
Since $\Vert\tilde{\phi}_{f}\Vert_{\tilde{\eta}}\Vert\tilde{\phi}_{i}%
\Vert_{\tilde{\eta}}=1$ in (\ref{gh}), we obtain with $\tilde{\beta}=1$ the
passage time $\tilde{\tau}=\pi/\tilde{\omega}$. Likewise we may evaluate
analogues to the other matrix elements computed in section II C. Our results
are summarized in table 2

\begin{center}%
\begin{tabular}
[c]{||l|l|l|l||}\hline
$\left\vert \psi_{i}\right\rangle $ & $\left\vert \psi_{f}\right\rangle $ &
$\tilde{U}(\tau,0)$ & $\tau$\\\hline\hline
$\left\vert \tilde{\Phi}_{i}\right\rangle $ & $\left\vert \tilde{\Phi}%
_{f}\right\rangle $ & $e^{-it\tilde{H}}$ & $\pi/\tilde{\omega}$\\\hline
$\left\vert \tilde{\phi}_{i}\right\rangle $ & $\left\vert \tilde{\phi}%
_{f}\right\rangle $ & $e^{-it\tilde{H}}$ & $\pi/\tilde{\omega}$\\\hline
$\left\vert \tilde{\phi}_{i}\right\rangle $ & $\left\vert \tilde{\Phi}%
_{f}\right\rangle $ & $e^{-it\tilde{H}}$ & $\pi/\tilde{\omega}$\\\hline
$\left\vert \tilde{\Phi}_{i}\right\rangle $ & $\left\vert \tilde{\phi}%
_{f}\right\rangle $ & $e^{-it\tilde{H}}$ & $\pi/\tilde{\omega}$\\\hline
\end{tabular}

\end{center}

\noindent{\small {Table 2: Passage times for various different initial and
final states evolved by means of a dissipative non-Hermitian time evolution
operator with real transition frequency. All matrix elements are computed with
the $\tilde{\eta}$-inner product and states are normalized with the $\eta
$-norm.}}

Thus somewhat surprisingly, despite the fact the matrix elements are somwhat
different, the normalization factors compensate for this and in all cases the
passage time results to $\tilde{\tau}=\pi/\tilde{\omega}$. This might suggest
that we should really attribute the possibility of tunable passage times to
the $\mathcal{PT}$-symmetry. However, we have not yet studied the possibility
when $\tilde{\omega}\notin\mathbb{R}$.

\subsection{Complex transition frequency}

Often it is not even possible to restrict the parameter range so nicely like
in the previous subsection as to ensure that $\tilde{\omega}\in\mathbb{R}$.
Instead considering the scenario of leaving this regime for the above example,
let us consider a different system, which does not even possess such a regime
and can be found for instance in \cite{IngR}%
\begin{equation}
\hat{H}=\left(
\begin{array}
[c]{cc}%
E+\varepsilon & 0\\
0 & E-\varepsilon
\end{array}
\right)  -\frac{i\lambda}{2}\left(
\begin{array}
[c]{cc}%
2\cos^{2}\phi & \sin2\phi\\
\sin2\phi & 2\sin^{2}\phi
\end{array}
\right)  ,~~ \label{com}%
\end{equation}
with $E,\varepsilon\in\mathbb{R}$ and $\lambda,\phi\in\mathbb{C}$. In
\cite{IngR} the special case $\lambda,\phi\in\mathbb{R}$ was treated for a
concrete physical setting. Here we keep these parameters completely generic.
The eigenvalues of $\hat{H}$ are
\begin{equation}
\hat{\varepsilon}_{\pm}=E\pm\frac{\hat{\omega}}{2}-i\frac{\lambda}{2},
\end{equation}
with energy gap $\hat{\omega}=\hat{\varepsilon}_{+}-\hat{\varepsilon}%
_{-}=\sqrt{4\varepsilon^{2}-\lambda^{2}-4i\varepsilon\lambda\cos2\phi}$. Now
we have lost the property of the transition frequency to be real. We
parameterize instead%
\begin{equation}
e^{-i\hat{\alpha}}=\frac{-i\lambda\sin2\phi}{2\varepsilon+\hat{\omega
}-i\lambda\cos2\phi}, \label{ahat}%
\end{equation}
such that the eigenvectors can simply be taken to be $|\hat{\Phi}_{\pm}%
\rangle=|\Phi_{\pm}\rangle_{\hat{\alpha}}$. Replacing now also in $\eta$, as
defined in (\ref{eta}), the parameter $\alpha$ by $\hat{\alpha}$ defines a new
operator $\hat{\eta}$. We employ this operator to construct the Hamiltonian
\begin{equation}
\hat{h}=\hat{\eta}\hat{H}\hat{\eta}^{-1}=\left(
\begin{array}
[c]{cc}%
E-i\frac{\lambda}{2} & -\frac{\hat{\omega}}{2}\\
-\frac{\hat{\omega}}{2} & E-i\frac{\lambda}{2}%
\end{array}
\right)  ,
\end{equation}
with eigenvectors $|\hat{\phi}_{\pm}\rangle:=|\phi_{\pm}\rangle$ as defined in
(\ref{ff}). We may now compute the passage time in a very similar fashion as
for the $\mathcal{PT}$-symmetric example, keeping however in mind that the
transition frequency $\hat{\omega}$ as well as the parameter $\hat{\alpha}$
are complex. A consequence of the latter is that $\hat{\eta}$ is no longer
Hermitian, that is $\hat{\eta}^{\dagger}\neq\hat{\eta}$. For convenience we
introduce the abbreviations $\hat{\omega}=\hat{\omega}_{r}+i\hat{\omega}_{i}$,
$\hat{\alpha}=\hat{\alpha}_{r}+i\hat{\alpha}_{i}$ and $\lambda=\lambda
_{r}+i\lambda_{i}$ for $\lambda_{i/r},\hat{\omega}_{i/r},\hat{\alpha}_{i/r}%
\in\mathbb{R}.$

\subsubsection{$\hat{\Phi}\rightarrow\hat{\Phi}$ via \^{U}}

It is straightforward to compute the square of the transition probability%
\begin{equation}
\left\vert \left\langle \hat{\phi}_{f}\right.  \!\left\vert e^{-ith}\hat{\phi
}_{i}\right\rangle \right\vert ^{2}=\frac{1}{2}e^{-\lambda_{r}t}\left[
\cosh(t\hat{\omega}_{i})-\cos(t\hat{\omega}_{r})\right]  , \label{tt}%
\end{equation}
when taking the initial and final states to be the orthonormal states as
defined in (\ref{fi}), with $|\phi_{\pm}\rangle\rightarrow|\hat{\phi}_{\pm
}\rangle.$ There are now various possibilities we can equate this to and
subsequently compute a passage time $\tau$. However, choosing a generic
$0\leq\beta\leq1$ as for instance in equation (\ref{qb}) would lead to very
involved solutions, due to the transcendental nature of this equation.
Furthermore, since we are now dealing with a dissipative system the right hand
side of (\ref{tt}) only reaches a maximum $\beta^{\prime}<1$. Thus the choice
$\beta>\beta^{\prime}$ would lead to unphysical complex passage times. In
order to keep matters simple, we could make a natural choice and just take
this maximum of the right hand side of (\ref{tt}), i.e. $\beta=\beta^{\prime}%
$. However, due to the transcendental nature of this equation, there is no
elegant analytic solution for $\tau$ to this equation. Instead we try to make
this solution to be as closely related to previously computed expressions as
possible and choose the right hand side such that the passage time becomes
$\tau=\pi/\hat{\omega}_{r}$ leading to a particular $\check{\beta}\leq
\beta^{\prime}$. We note that this choice does not lead to a loss of
generality with regard to the main aim of our investigation, which is to seek
passage times which can be made arbitrarily small. Taking any other value
below the maximum will simply lead to another definite value for $\tau$, when
keeping $\hat{\omega}$ constant.

\subsubsection{$\hat{\Phi}\rightarrow\hat{\Phi}$ via \^{u}}

As in all previous scenarios we take the special case $\hat{\Phi}%
\rightarrow\hat{\Phi}$ via \^{U} as a benchmark for the choice of the constant
$\beta=\check{\beta}$, since it can be dealt with analytically. Let us now
compute%
\begin{equation}
\left\vert \left\langle \hat{\phi}_{f}\right.  \!\left\vert e^{-it\hat{H}}%
\hat{\phi}_{i}\right\rangle _{\hat{\eta}}\right\vert ^{2}=\frac{\cosh
(t\hat{\omega}_{i})-\cos(2\hat{\alpha}_{r}-t\hat{\omega}_{r})}{2e^{\lambda
_{r}t}\cos\hat{\alpha}\cos\hat{\alpha}^{\ast}}.
\end{equation}
With
\begin{equation}
\left\Vert \hat{\phi}_{f}\right\Vert _{\hat{\eta}}^{2}\left\Vert \hat{\phi
}_{i}\right\Vert _{\hat{\eta}}^{2}=\frac{\cosh^{2}\hat{\alpha}_{i}}{\cos
\hat{\alpha}\cos\hat{\alpha}^{\ast}}%
\end{equation}
and the choice of $\beta=\check{\beta}$ as discussed in the previous section,
namely taking it to be the right hand side of (\ref{tt}) with $t\rightarrow
\pi/\hat{\omega}_{r}$, the quantum brachistochrone problem amounts to solving
\begin{equation}
\frac{\cosh(t\hat{\omega}_{i})-\cos(2\hat{\alpha}_{r}-t\hat{\omega}_{r}%
)}{\cosh^{2}\hat{\alpha}_{i}\left(  1+\cosh(\pi\hat{\omega}_{i}/\hat{\omega
}_{r})\right)  }=e^{\lambda_{r}(t-\pi/\hat{\omega}_{r})} \label{trans}%
\end{equation}
in this case. Since this is a transcendental equation, we can not solve it in
complete generality and we are therefore content to discuss some numerical
solutions. For this purpose we can solve the equation of the transition
frequency for the coupling constant $\lambda=-2i\varepsilon\cos\phi
+\sqrt{4\varepsilon^{2}\sin^{2}2\phi-\omega^{2}}$ and express it as a function
of $\phi,\varepsilon$ and $\omega$. Since we want to keep $\omega$ constant,
we investigate (\ref{trans}) as a function of time $t$ by varying $\phi$ and
$\lambda$ or $\varepsilon$ and $\lambda$ see figure 1 or figure 2, respectively.

The analysis of equation (\ref{trans}), as depicted in figures 1 and 2
demonstrates that it is possible to find tunable passage times for Hamiltonian
systems of the type (\ref{com}). The precise dependence of the system on the
parameters is rather involved in this case, but our analysis demonstrates that
it is possible to approach $\tau\approx0$. Similar conclusions can be drawn
when changing $|\phi_{f}\rangle$ to $|\Phi_{f}\rangle$ or $|\phi_{i}\rangle$
to $|\Phi_{i}\rangle$, respectively.

\noindent\epsfig{file=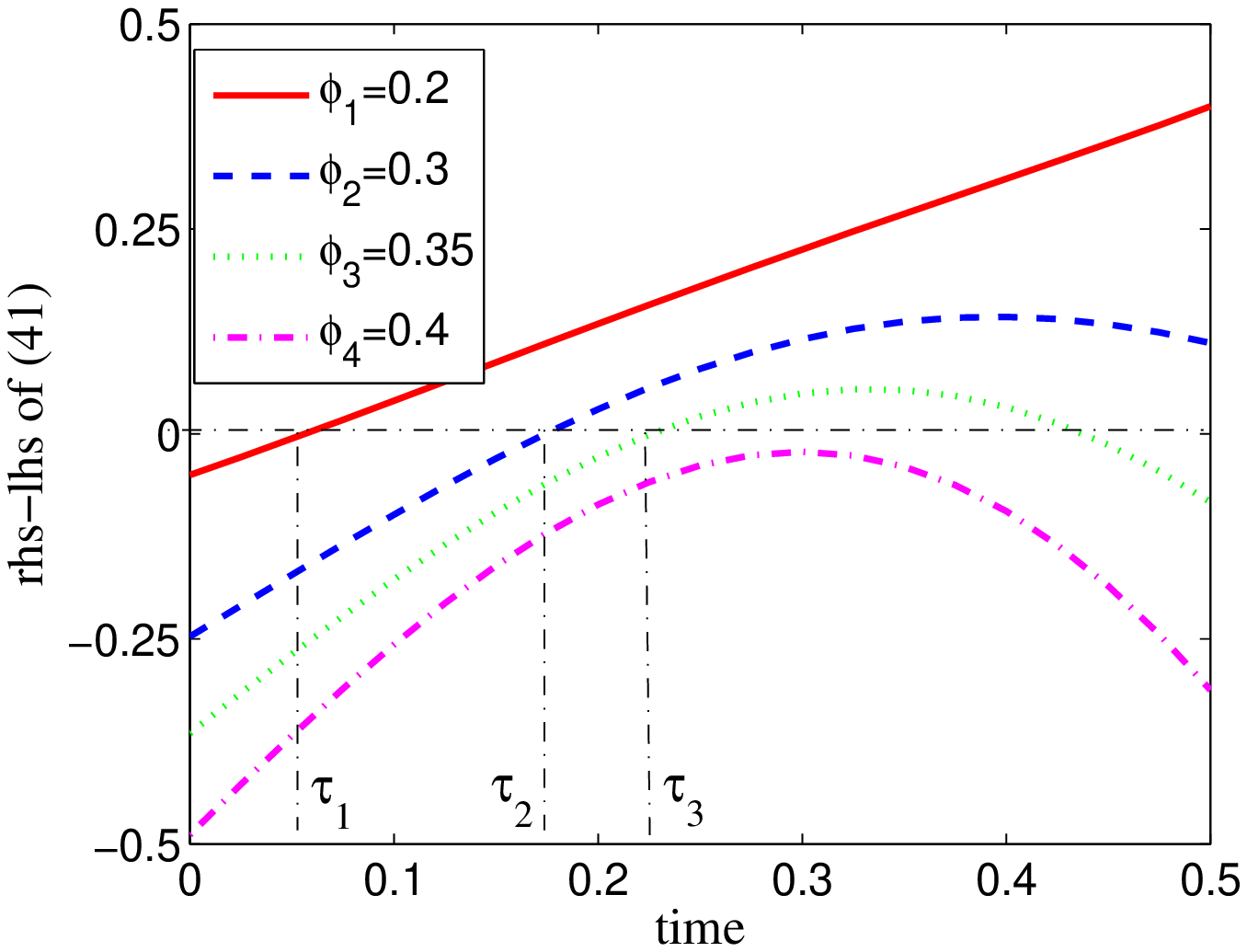,width=9.1cm}

{\small Figure 1: Right hand side(rhs) minus the left hand side(lhs) of
equation (\ref{trans}) as a function of time $t$ with fixed values of
$\varepsilon=2.5,E=2,\omega=4.87-i3.31$ for $\phi_{1}=0.2,\lambda_{1}=3.5$ ;
$\phi_{2}=0.3,\lambda_{2}=3.73+i0.2$ ; $\phi_{3}=0.35,\lambda_{3}=3.87+i0.34$
; $\phi_{4}=0.4,\lambda_{4}=4.02+i0.52$. The corresponding passage times
$\tau$ are simply taken to be the smallest values of t for which the
rhs-lhs=0.}

\noindent\epsfig{file=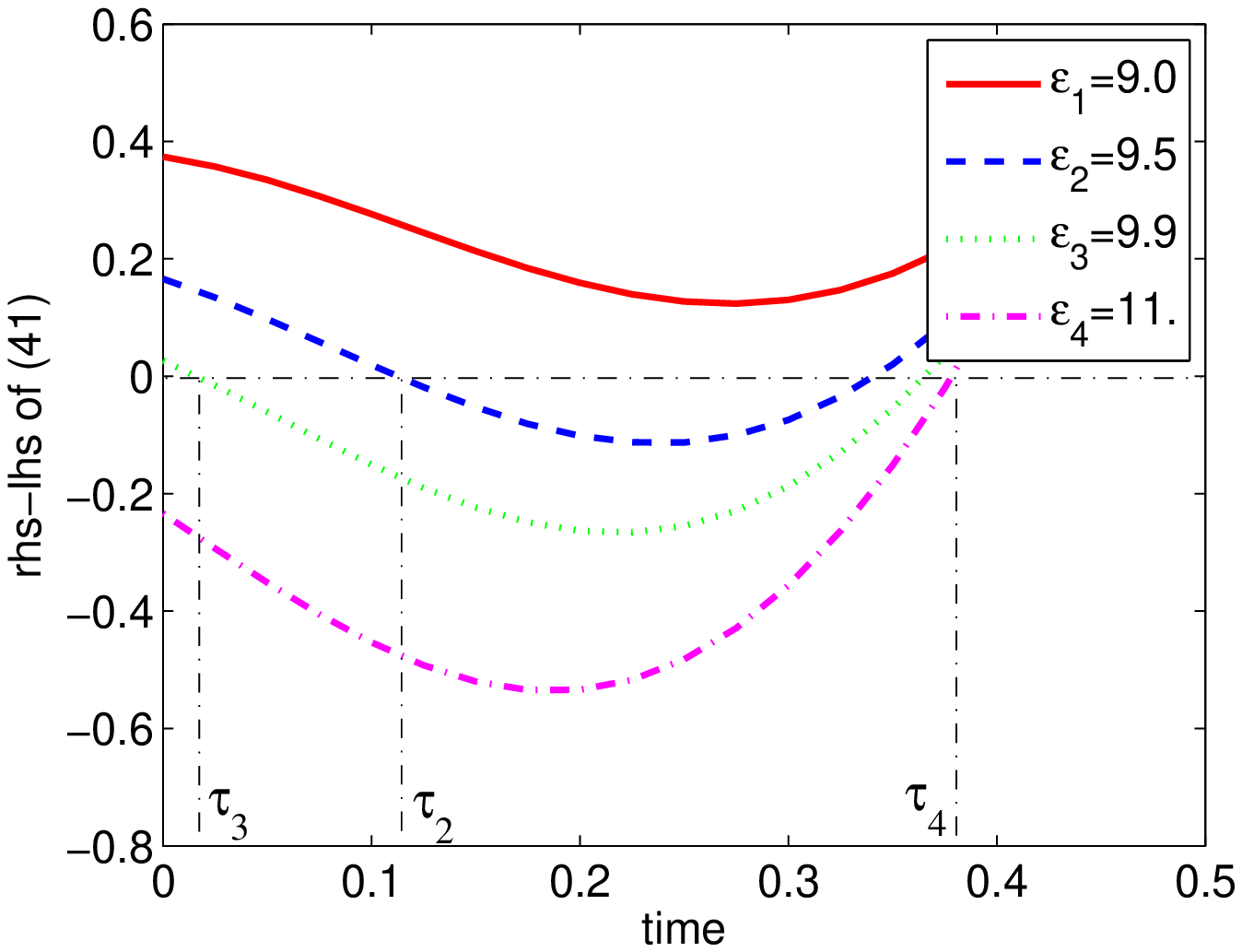,width=9.1cm}

{\small Figure 2: Right hand side minus the left hand side of equation
(\ref{trans}) as a function of time $t$ with fixed values of }${\small \phi}%
${\small $=0.2,E=2,\omega=6.99-i0.46$ for }${\small \varepsilon}$%
{\small $_{1}=9.0,\lambda_{1}=1.85-i14.84$ ; }${\small \varepsilon}%
${\small $_{2}=9.5,\lambda_{2}=2.72-i16.32$ ; }${\small \varepsilon}%
${\small $_{3}=9.9,\lambda_{3}=3.41-i17.29$ ; }${\small \varepsilon}%
${\small $_{4}=11.0,\lambda_{4}=5.01-i19.62$.}

\section{Conclusion}

In \cite{faster} the authors have extended the formulation of the quantum
brachistochrone problem by allowing that the time-evolution operator may be
associated to non-Hermitian Hamiltonians which are $\mathcal{PT}$-symmetric,
that is those with real eigenvalues. Here we did not insist on the
$\mathcal{PT}$-symmetry of the Hamiltonian, but allowed this symmetry to be
completely broken. In order to take the most extreme case, we did not just
spontaneously break the $\mathcal{PT}$-symmetry for the wavefunction, which
would result in complex conjugate pairs for the energy eigenfunction, but we
allowed in addition that the $\mathcal{PT}$-symmetry is also broken for the
Hamiltonian. Thus we have considered an effective Hamiltonian whose energy
eigenvalues have a negative imaginary part, such that it is associated to
dissipative systems. We found the same intriguing feature as observed in
\cite{faster} for the quantum brachistochrone problem for $\mathcal{PT}%
$-symmetric non-Hermitian Hamiltonians, namely that the passage time can be
made arbitrarily small also for non-Hermitian Hamiltonians associated to these
type of systems. Our observations suggest that this type of phenomenon may
occur when one projects between orthonormal states, which are not eigenstates
of the non-Hermitian Hamiltonian associated to the time-evolution operator,
irrespective of whether this Hamiltonian is $\mathcal{PT}$-symmetric or not.

Clearly there are various open questions to be answered. First of all, it
would be highly desirable to have a more formal and generic proof for this
phenomena, rather than case-by-case studies. This holds for the $\mathcal{PT}%
$-symmetric case treated in \cite{faster} as well as for the case presented
here. In addition, one could make the above considerations more involved by
allowing more complicated time-dependences rather than the simple stepfunction
and study other possibilities in (\ref{h1}). For such more realistic scenarios
we may have to resort to a perturbative treatment using (\ref{12}).

\medskip

\noindent\textbf{Acknowledgments}. Discussions with D.C. Brody, C. Figueira de
Morisson Faria, H. Jones, M. Khokhlova and I. Rotter are gratefully
acknowledged. P.E.G.A. is supported by a City University research studentship.

\end{document}